\documentstyle[12pt,epsf]{article}

\newcommand\be{\begin{equation}}
\newcommand\ee{\end{equation}}
\newcommand\bea{\begin{eqnarray}}
\newcommand\eea{\end{eqnarray}}

\newcommand\lae{\stackrel{<}{\sim}}
\newcommand\gae{\stackrel{>}{\sim}}

\newcommand\iic{\ \ ,}
\newcommand\iip{\ \ .}

\newcommand\tr{{\rm Tr}}
\newcommand\gev{{\rm GeV}}

\begin{document}

\begin{titlepage}
\def\thepage {}     

\title{Limits on Flavor-Universal Colorons\thanks{Talk given on July
7, 1996 at the Workshop on New Directions in High-Energy Physics
(Snowmass 96).}}
\author{Elizabeth H. Simmons\thanks{e-mail address:
simmons@bu.edu}\thanks{E.H.S. acknowledges the hospitality of the
Aspen Center for Physics and the Fermilab Summer Visitors Program and
the support of the NSF CAREER and DOE OJI programs. {\em This work was
supported in part by the National Science Foundation under grant
PHY-95-1249 and by the Department of Energy under grant
DE-FG02-91ER40676.}  } \\
Department of Physics, Boston University \\
590 Commonwealth Ave., Boston  MA  02215}

\date{\today}
\maketitle

\bigskip
\begin{picture}(0,0)(0,0)
\put(295,250){BUHEP-96-27}
\put(295,235){hep-ph/9608349}
\end{picture}
\vspace{24pt}

\begin{abstract}
A flavor-universal extension of the strong interactions was recently
proposed in response to the apparent excess of high-$E_T$ jets in the
inclusive jet spectrum measured at the Tevatron.  The color octet of
massive gauge bosons (`colorons') that is present in the low-energy
spectrum of the model's Higgs phase is studied here.  Experimental
constraints already imply that the coloron mass must exceed 870-1000
GeV.  The import of recent Tevatron data and the prospective input
from future experiments are also mentioned.
  
\pagestyle{empty}
\end{abstract}

\end{titlepage}



\section{Introduction}

A flavor-universal coloron model \cite{newint} was recently proposed
to explain the apparent excess of high-$E_T$ jets in the inclusive jet
spectrum measured by CDF \cite{CDFexc}.  This model is a
flavor-universal variant of the coloron model of Hill and Parke
\cite{topglu} which can accommodate the jet excess without
contradicting other experimental data.  It involves a minimal
extension of the standard description of the strong interactions,
including the addition of one gauge interaction and a scalar
multiplet, but no new fermions.  As such, it serves as a useful
baseline with which to compare both the data and other candidate
explanations of the jet excess \cite{pdfs}.  Furthermore, the extended
strong interactions can be grafted onto the standard one-Higgs-doublet
model of electroweak physics, yielding a simple, complete, and
renormalizable theory.

Here, we briefly describe the phenomenology of the Higgs phase of the
model (for a fuller discussion see ref. \cite{colph}).  We discuss the
limits which current data places on the colorons and then indicate how
future measurements may be of use.

\section{The model}
\label{sec:model}
\setcounter{equation}{0}

In the flavor-universal coloron model \cite{newint}, the strong gauge
group is extended to $SU(3)_1 \times SU(3)_2$.  The gauge couplings
are, respectively, $\xi_1$ and $\xi_2$ with $\xi_1 \ll \xi_2$.  Each
quark transforms as a (1,3) under this extended strong gauge group.

The model also includes a scalar boson $\Phi$ transforming as a
$(3,\bar 3)$ under the two $SU(3)$ groups.  The most
general\footnote{As noted in \cite{newint} this model can be grafted
onto the standard one-doublet Higgs model.  In this case, the most
general renormalizable potential for $\Phi$ and the Higgs doublet
$\phi$ also includes the term $\lambda_3 \phi^\dagger\phi
\tr(\Phi^\dagger\Phi)$.  For a range of $\lambda$'s and parameters in
the Higgs potential, the vacuum will break the two $SU(3)$ groups to
QCD and also break the electroweak symmetry as required.} potential for
$\Phi$ is
\be
U(\Phi) = \lambda_1 \tr\left(\Phi\Phi^\dagger - f^2 {\rm I}\right)^2 +
\lambda_2 \tr\left(\Phi\Phi^\dagger - \frac 13 {\rm I} \left(\tr
\Phi\Phi^\dagger\right)\right)^2
\ee
where the overall constant has been adjusted so that the minimum of
$U$ is zero.  For $\lambda_1,\,\lambda_2,\,f^2 > 0$ the scalar
develops a vacuum expectation value $\langle\Phi\rangle = {\rm
diag}(f,f,f)$ which breaks the two strong groups to their diagonal
subgroup.  We identify this unbroken subgroup with QCD.

The original gauge bosons mix to form an octet of massless gluons and
an octet of massive colorons.  The gluons interact with quarks through
a conventional QCD coupling with strength $g_3$.  The colorons
$(C^{\mu a})$ interact with quarks through a new QCD-like coupling
\be
{\cal L} = - g_3  \cot\theta J^a_\mu C^{\mu a} \iic
\ee
where  $J^a_\mu$ is the color current
\be
\sum_f {\bar q}_f \gamma_\mu \frac{\lambda^a}{2}q_f \iip
\ee 
and $\cot\theta = \xi_2/\xi_1\, $.  Note that we expect $\cot\theta
>1$.  In terms of the QCD coupling, the gauge boson mixing angle and
the scalar vacuum expectation value, the mass of the colorons is
\be
M_c = \left( \frac{g_3}{\sin\theta \cos\theta} \right) f \iip
\ee
The colorons decay to all sufficiently light quarks; assuming
there are $n$ flavors lighter than $M_c/2$, the decay width is
\be
\Gamma_c \approx \frac n6 \alpha_s \cot^2\theta\, M_c
\label{eqwid}
\ee
where $\alpha_s \equiv g_3^2/4\pi$. We take the top quark mass to be
175 GeV so that n = 5 for $M_c \lae 350$ GeV and n = 6 otherwise.

\section{Existing limits on colorons}
\label{sec:lim-now}
\setcounter{equation}{0}

A sufficiently light coloron would be visible in direct production at
the Tevatron.  Indeed, the CDF Collaboration has searched for new
particles decaying to dijets and reported \cite{CDFdij} an upper limit
on the incoherent production of such states.  Accordingly, as
discussed in \cite{colph}, we calculated $\sigma \cdot B$ for colorons
with various values of $M_c$ and $\cot\theta$.  We followed the
example of CDF in using CTEQ structure functions and in requiring
$\vert\eta\vert < 2$ and $\vert\cos\theta^*\vert < 2/3$.  For
$\cot^2\theta < 2$, the coloron's half-width falls within the dijet
mass resolution of 10\%; for larger $\cot\theta$ we counted only the
portion of the signal that falls within a bin centered on the coloron
mass and with a width equal to the resolution.  Values of $M_c$ and
$\cot\theta$ which yield a theoretical prediction that exceeds the CDF
upper limit are deemed to be excluded at 95\% c.l.  \cite{CDFdij}. We
find that for $\cot\theta = 1$, the range $200\ {\rm GeV} < M_c < 870\
{\rm GeV}$ is excluded; at $\cot\theta = 1.5$, the upper limit of the
excluded region rises to roughly 950 GeV; at $\cot\theta = 2$, it
rises to roughly 1 TeV.  As the coloron width grows like
$\cot^2\theta$, going to higher values of $\cot\theta$ does not
appreciably increase the upper limit of the excluded range of masses
beyond 1 TeV.

\begin{figure}[htb]
\vspace{-3.5cm}
\epsfxsize 10cm \centerline{\epsffile{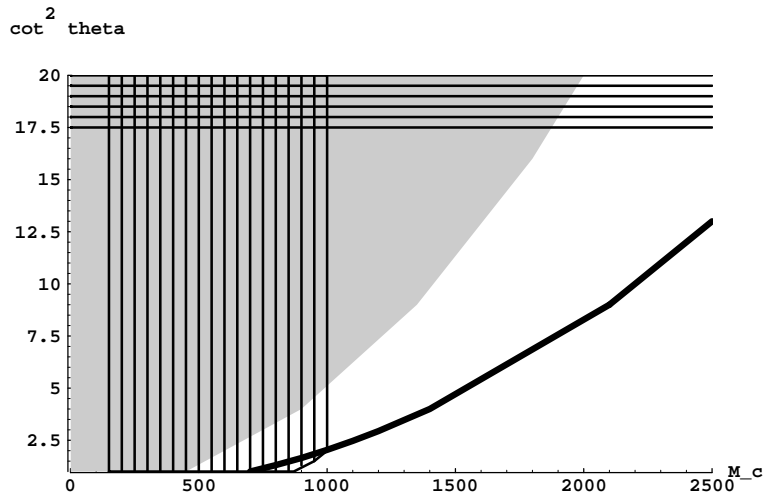}}
\vspace{-3.5cm}
\caption[allcurlim]{Current limits on the coloron parameter space: mass ($M_c$)
vs. mixing parameter ($\cot^2\theta$).  The shaded region is excluded
by the weak-interaction $\rho$ parameter \cite{newint} as in equation
\ref{rhoweq}.  The vertically-hatched polygon is excluded by searches
for new particles decaying to dijets \cite{CDFdij,ua1}. The
horizontally-hatched region at large $\cot^2\theta$ lies outside the
Higgs phase of the model.  The dark line is the curve $M_c /
\cot\theta = 700$ GeV for reference.} 
\label{allcurlim}
\end{figure}

To extend the excluded range of coloron masses to values below those
probed by CDF, we note two things.  First, $\sigma$ increases as
$\cot\theta$ does, so that exclusion of $\cot\theta = 1$ for a given
$M_c$ implies exclusion of all higher values of $\cot\theta$ at that
$M_c$.  Second, $\sigma \cdot B$ is the same for a coloron with
$\cot\theta = 1$ as for an axigluon \cite{axig} of identical mass
\cite{colph}.  Axigluons with masses between 150 and 310 GeV have
already been excluded by UA1's analysis \cite{ua1} of incoherent
axigluon production; by extension, colorons in this mass range with
$\cot\theta \geq 1$ are also excluded.  The combined excluded ranges
of $M_c$ are
\bea
150 \gev < &M_c& < 870 \gev \ \ \ \ \ \ \cot\theta = 1\nonumber \\
150 \gev < &M_c& < 950 \gev \ \ \ \ \ \ \cot\theta = 1.5 \\
150 \gev < &M_c& < 1000 \gev \ \ \ \ \ \cot\theta \gae 2\iip\nonumber
\eea
They are summarized by the shaded region of figure \ref{allcurlim}

\medskip

Because the colorons couple to all flavors of quarks, they should also
affect the sample of b-tagged dijets observed at Tevatron experiments.
As discussed in ref. \cite{colph}, however, the limit on colorons from
b-tagged dijets will probably be weaker than that from the full dijet
sample.  This contrasts with the case of topgluons, which can be more
strongly constrained by the b-tagged dijet sample because they decay
almost exclusively to third-generation quarks.

\medskip

An additional limit on the coloron mixing angle may be derived from
constraints on the size of the weak-interaction $\rho$-parameter.
Coloron exchange across virtual quark loops contributes to
$\Delta\rho$ through the isospin-splitting provided by the difference
between the masses of the top and bottom quarks.  Limits on this type
of correction \cite{cdt} imply that \cite{newint}
\be
\frac{M_c}{\cot\theta} \gae 450 {\rm GeV} \iip
\label{rhoweq}
\ee
This excludes the hatched region of the $\cot^2\theta$ -- $M_c$ plane
shown in figure \ref{allcurlim}.  Note that this excludes an area of
small $M_c$ that the dijet limits did not probe, as well as an area at
larger $M_c$ and large $\cot\theta$.

\medskip

Finally, we mention a theoretical limit on the coloron parameter
space.  While the model assumes $\cot\theta > 1$, the value of
$\cot\theta$ cannot be arbitrarily large if the model is to be in the
Higgs phase at low energies.  Writing the low-energy interaction among
quarks that results from coloron exchange as a four-fermion
interaction
\be
{\cal L}_{4f} = - \frac{g_3^2 \cot^2\theta}{M_c^2} J^a_\mu J^{a \mu}
\label{fourff}
\ee
we use the NJL approximation to estimate the critical value of
$\cot^2\theta$ as\footnote{In more conventional notation (see e.g.
\cite{topglu}), one would write the coefficient of the four-fermion
operator as $-(4\pi\kappa/M^2)$ and find $\kappa_{crit} = 2 \pi/ 3$.}
\be
(\cot^2\theta)_ {crit} = \frac{2\pi}{3 \alpha_s}  \approx 17.5
\ee
This puts an upper limit on the $\cot^2\theta$ axis of the coloron's 
parameter space, as indicated in figure \ref{allcurlim}.

\section{Upcoming limits from Tevatron data}
\label{sec:lim-when}
\setcounter{equation}{0}

\begin{figure}[htb]
\vspace{-3.5cm}
\epsfxsize 10cm \centerline{\epsffile{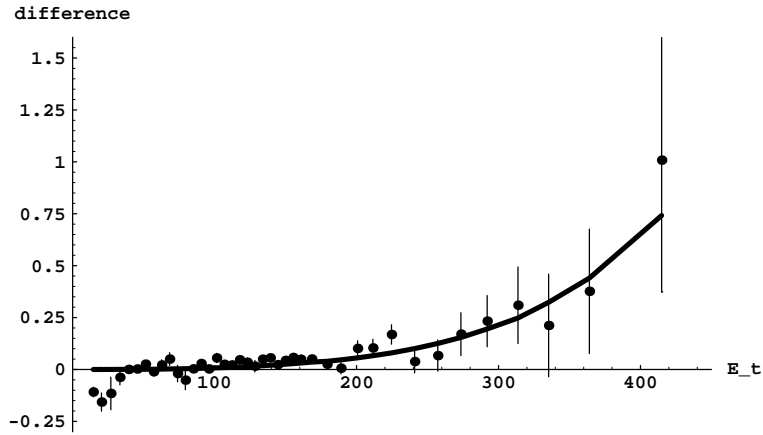}}
\vspace{-3.5cm}
\caption[differ]{Difference plot ((data - theory)/theory) for the
inclusive jet cross-section ${1\over{\Delta\eta}} \int
(d^2\sigma/d\eta\, dE_T) d\eta $ as a function of transverse jet
energy $E_t$, where the pseudorapidity $\eta$ of the jet falls in the
range $0.1 \leq \vert\eta\vert \leq 0.7$. Dots with (statistical)
error bars are the recently published CDF data \protect\cite{CDFexc}.
The solid curve shows the LO prediction of QCD plus the contact
interaction approximation to coloron exchange of equation
(\protect\ref{fourff}) with $M_C/\cot\theta = 700$ GeV.  Following
CDF, we employed the MRSD0' structure functions \protect\cite{mrs_pak}
and normalized the curves to the data in the region where the
effect of the contact interactions is small (here this region is $45 <
E_T < 95$ GeV).}
\label{diffc}
\end{figure}

Both the inclusive jet spectrum ($d\sigma/ d E_T$) and the dijet
invariant mass spectrum ($d\sigma / d M_{jj}$) measured in CDF's run
IA and IB data \cite{CDFexc} appear to show excesses at high energy
end of the spectrum.  The dijet limits we derived earlier imply that
the coloron is heavy enough that it would not be directly produced in
the existing Tevatron data.  Therefore, it is useful to start studying
the data in terms of the four-fermion approximation (\ref{fourff}) to
coloron exchange. Comparison with the run IA CDF inclusive jet
spectrum already \cite{newint} indicates that $M_c / \cot\theta = 700$
GeV is not obviously ruled out, as figure \ref{diffc}
illustrates. Figure
\ref{allcurlim} indicates where the curve $M_c / \cot\theta = 700$ GeV
falls relative to the limits on the parameter space discussed earlier. 
Detailed analysis including both systematic and statistical errors
should be able to determine a lower bound on $M_c / \cot\theta$.

For colorons weighing a little more than a TeV -- those that are just
above the current dijet mass bound -- it is more appropriate to use
the cross-sections for full coloron exchange \cite{colph} when making
comparisons with the data.  Such colorons are light enough that their
inclusion yields a cross-section of noticeably different shape than
the four-fermion approximation would give (see figure 3).  Once the
full coloron-exchange cross-sections are employed, the mass and mixing
angle of the coloron may be varied independently.  In particular, one
may study the effects of light colorons with small values of
$\cot^2\theta$.  This can expand the range of accessible parameter
space beyond what one would have reached by using the four-fermion
approximation.

\begin{figure}[htb]
\vspace{-3.5cm}
\epsfxsize 10cm \centerline{\epsffile{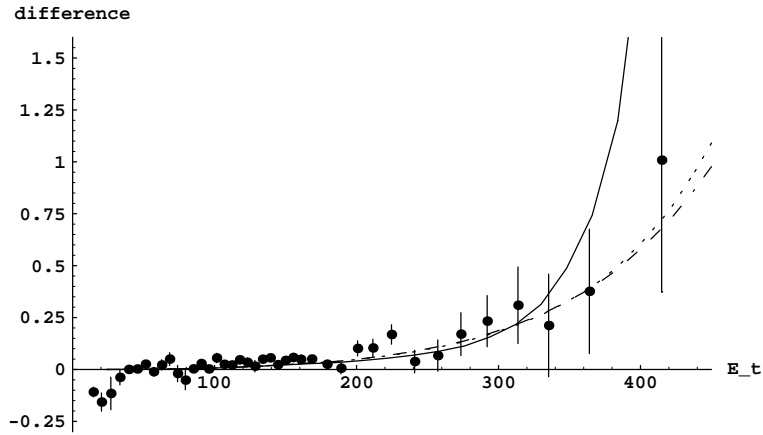}}
\vspace{-3.5cm}
\caption[shape]{Difference plot for $d\sigma / d E_T$ (see 
figure 3) showing the effects of colorons of different masses when the
ratio $M_c / \cot\theta$ is fixed at 700 GeV.  Here full one-coloron
exchange is included, rather than the contact interaction
approximation.  The solid curve is for a light coloron: $M_c = 1050$
GeV, $\cot\theta = 1.5$.  The dotted and dashed curves correspond to much
heavier colorons ($M_c = 1750$ GeV and 2000 GeV) with correspondingly
larger values of $\cot\theta$ (2.5 and 3.0).  The cross-section for
the heavier colorons is well-approximated by the contact interaction 
approximation at Tevatron energies; the cross-section for the lighter coloron
is not.}
\label{shape}
\end{figure}

\medskip
 
Another means of determining what kind of new strong interaction is
being detected is measurement of the dijet angular distribution.  Some
new interactions would produce dijet angular distributions like that
of QCD; others predict distributions of different shape.  In terms of
the angular variable $\chi$ %
\be
\chi = \frac{1 + \vert\cos\theta^*\vert}{1 - \vert\cos\theta^*\vert}
\ee
QCD-like jet distributions appear rather flat while those which are
more isotropic in $\cos\theta^*$ peak at low $\chi$ (recall that
$\theta^*$ is the angle between the proton and jet directions).  The
ratio $R_\chi$
\be
R_\chi \equiv \frac{N_{events}, 1.0 < \chi < 2.5}{N_{events}, 2.5 <
\chi < 5.0}
\ee
then captures the shape of the distribution for a given sample of
events, e.g. at a particular dijet invariant mass.

The CDF Collaboration has made a preliminary analysis of the dijet
angular information in terms of $R_\chi$ at several values of dijet
invariant mass \cite{rmh}.  The preliminary data appears to be
consistent either with QCD or with QCD plus a color-octet four-fermion
interaction like (\ref{fourff}) for $M_c/\cot\theta = 700$ GeV.  Our
calculation of $R_\chi$ including a propagating coloron gives results
consistent with these.  It appears that the measured angular
distribution can allow the presence of a coloron and can help put a
lower bound on $M_c/\cot\theta$.

\section{Conclusions and Prospects}
\label{sec:concl}
\setcounter{equation}{0}

The flavor-universal coloron model can accommodate an excess at the
high-$E_t$ end of the inclusive jet spectrum at Tevatron energies
without contradicting other data.  Previous measurements of the
weak-interaction $\rho$ parameter and searches for new particles
decaying to dijets imply that the coloron must have a mass of at least
870 GeV.  Measurements of jet spectra and angular distributions from
runs IA and IB at the Tevatron, from future Tevatron runs, and
eventually from the LHC will shed further light on the model.

This model would be even more interesting if it could also shed light
on the origins of electroweak and flavor symmetry breaking.  The
minimal form described here has little connection to electroweak
physics and none to flavor physics.  However, it appears possible to
include the extended strong interactions within a framework that
addresses both issues.  The simplest possibility\footnote{The author
acknowledges the invaluable input of R.S. Chivukula and N. Evans on
this topic.} is to build a variant of topcolor-assisted technicolor
\cite{tc2} in which electroweak symmetry breaking arises largely from
technicolor, the strong interactions are as in the flavor-universal
coloron model, and a pair of flavor-discriminating $U(1)$ interactions
cause formation of a top condensate (and therefore a large top quark
mass).  Such a model would have low-energy jet phenomenology
resembling that of the flavor-universal coloron model and potentially
smaller FCNC effects involving bottom quarks, but would also have a
large number of relatively light scalar bound states made of the light
quarks.



\begin{thebibliography}{99}
\frenchspacing

\bibitem{newint} ``New Strong Interactions at the Tevatron?'', R.S.
Chivukula, A.G. Cohen, and E.H. Simmons, hep-ph/9603311, to appear in
Physics Letters B (1996).

\bibitem{CDFexc} ``Inclusive Jet Cross Section in $\bar p p$
Collisions at $\sqrt{s} = 1.8$ TeV'', CDF Collaboration, F.~Abe {\it
et al.}, FERMILAB-PUB-96/020-E, hep-ex/9601008.

\bibitem{topglu} C.T. Hill Phys. Lett. {\bf B266} (1991) 419;
C.T. Hill and S.J. Parke, Phys. Rev. {\bf D49} (1994) 4454.

\bibitem{pdfs} ``Improved Parton Distributions from Global Analysis if
Recent Deep Inelastic Scattering and Inclusive Jet Data,'' H.L. Lai et
al., MSUHEP-60426, CTEQ-604,(1996), hep-ph/9606399; ``Large Transverse Momentum
Jet Production and DIS Distributions of the Proton,'' M. Klasen and G. Kramer,
DESY 96-077 (1996). hep-ph/9605210; ``Hadrophyllic $Z^\prime$: A Bridge from
LEP1, SLC and CDF to LEP2 anomalies'', P. Chiapetta,  J. Layssac, F.M. Renard,
and C. Versegnassi, hep-ph/9601306; ``$R_b$, $R_c$ and Jet Distributions
at the Tevatron in a Model with an Extra Vector Boson'', G. Altarelli,
N. Di Bartolomeo, F. Feruglio, R. Gatto, and M.L. Mangano,
hep-ph/9601324; ``Quark Resonances and high-$E_T$ Jets'', M. Bander,
hep-ph/9602330; ``High $E_T$ Jets at $p\bar p$ Collisions and Triple
Gauge Vertex'', B.A. Arbuzov, hep-ph/9602416; ``Light Gluinos and Jet Production
in $\bar p p$ Collisions'', Z. Bern, A.K. Grant, and A.G. Morgan,
hep-ph/9606466.

\bibitem{colph} ``Coloron Phenomenology'', E.H. Simmons, (1996).  BUHEP-96-24.
hep-ph/9608269.

\bibitem{CDFdij} CDF Collaboration (F. Abe et al.)
Phys. Rev. Lett. {\bf 74} (1995) 3538.  hep-ex/9501001.

\bibitem{axig} J. Pati and A. Salam, Phys. Lett. {\bf 58B} (1975) 333;
J. Preskill, Nucl. Phys. {\bf B177} (1981) 21; L. Hall and A. Nelson,
Phys. Lett. {\bf 153B} (1985) 430; P.H. Frampton and S.L. Glashow,
Phys. Lett. {\bf B190} (1987) 157 and Phys. Rev. Lett. {\bf 58} (1987)
2168; J. Bagger, C. Schmidt, and S. King, Phys. Rev. 
{\bf D37} (1988) 1188. 

\bibitem{ua1} C. Albajar et al. (UA1 Collaboration), Phys. Lett. {\bf
B209} (1988) 127.

\bibitem{cdt} R.S. Chivukula, B.A. Dobrescu, and J. Terning, Phys.
Lett. {\bf B353} (1995) 289.

\bibitem{mrs_pak} A.D. Martin, R.G. Roberts and W.J. Stirling,
Phys. Lett. {\bf B306} (1993) 145.

\bibitem{rmh} R.M. Harris, private communication.  See preliminary CDF
results on the World Wide Web at
http://www-cdf.fnal.gov/physics/new/qcd/qcd\_plots /twojet/public/dijet\_new\_physics.html\ \ \ .  

\bibitem{tc2} C.T. Hill, Phys. Lett. {\bf B345} 483 (1995).


\end{thebibliography}
\end{document}